\documentclass[aps,pre,twocolumn,superscriptaddress,showkeys,longbibliography,floatfix]{revtex4-1}

\usepackage{amsmath,amssymb,amsfonts,amsthm,enumerate,multirow,mathtools}
\usepackage{color}
\usepackage{siunitx}
\usepackage[hidelinks]{hyperref}
\usepackage{placeins} 
\usepackage{booktabs}
\usepackage{algorithm,algpseudocode}
\usepackage{natbib}
\usepackage{subcaption}

\newcommand{\pmat}[1]{\begin{bmatrix}#1\end{bmatrix}}

\newcommand{\Transp}{\mathsf{T}}

\newcommand{\red}[1]{{\color{black}#1}}

\begin{document}
\newcommand{\coverTitle}{Tomographic Reconstruction of Triaxial Strain Fields from Bragg-Edge Neutron Imaging}
\newcommand{\coverAuthors}{J.N. Hendriks, A.W.T. Gregg, R.R. Jackson, C.M. Wensrich, A. Wills, A.S. Tremsin, T. Shinohara, V. Luzin, O. Kirstein.}
\newcommand{\coverStatus}{Accepted for publication.}

\begin{titlepage}
    \begin{center}
        {\large \em Technical report}
        
        \vspace*{2.5cm}
        %
        {\Huge \bfseries \coverTitle  \\[0.4cm]}
        
        %
        {\Large \coverAuthors \\[2cm]}
        
        \renewcommand\labelitemi{\color{red}\large$\bullet$}
        \begin{itemize}
            \item {\Large \textbf{Please cite this version:}} \\[0.4cm]
            \large
            \coverAuthors. \coverTitle. \textit{Phys. Rev. Materials, 3:113803, Nov 2019. doi: 10.1103/PhysRevMaterials.3.113803.}  
        \end{itemize}
        
        \vfill

        
        \vfill
        \vspace{50mm}
    \end{center}
\end{titlepage}
\newpage
\thispagestyle{empty}
\newpage

\title{Tomographic Reconstruction of Triaxial Strain Fields from Bragg-Edge Neutron Imaging}

\author{J.N. Hendriks}
\email[]{johannes.hendriks@uon.edu.au}
\affiliation{School of Engineering, The University of Newcastle, Callaghan NSW 2308, Australia}

\author{A.W.T. Gregg}
\affiliation{School of Engineering, The University of Newcastle, Callaghan NSW 2308, Australia}

\author{R.R. Jackson}
\affiliation{School of Engineering, The University of Newcastle, Callaghan NSW 2308, Australia}

\author{C.M. Wensrich}
\affiliation{School of Engineering, The University of Newcastle, Callaghan NSW 2308, Australia}

\author{A. Wills}
\affiliation{School of Engineering, The University of Newcastle, Callaghan NSW 2308, Australia}

\author{A.S. Tremsin}
\affiliation{Space Sciences Laboratory, University of California, Berkeley CA 94720, USA}

\author{T. Shinohara}
\affiliation{Materials and Life Sciences Facility, Japan Proton Accelerator Research Complex, Tokai-mura, Ibaraki 319-1195, Japan}

\author{V. Luzin}
\affiliation{ACNS, Australian Nuclear Science and Technology Organisation (ANSTO), Kirrawee NSW 2232, Australia}

\author{O. Kirstein}
\affiliation{School of Engineering, The University of Newcastle, Callaghan NSW 2308, Australia}
\affiliation{European Spallation Source, Lund 223 63, Sweden}

\date{\today}

\begin{abstract}
This paper presents a proof-of-concept demonstration of triaxial strain tomography from Bragg-edge neutron imaging within a three-dimensional sample. Bragg-edge neutron transmission can provide high-resolution images of the average through thickness strain within a polycrystalline material.
This poses an associated rich tomography problem which seeks to reconstruct the full triaxial strain field from these images. 
The presented demonstration is an important step towards solving this problem, and towards a technique capable of studying the residual strain and stress within engineering components.
A Gaussian process based approach is used that ensures the reconstruction satisfies equilibrium and known boundary conditions.
This approach is demonstrated experimentally on a non-trivial steel sample with use of the RADEN instrument at the Japan Proton Accelerator Research Complex. 
Validation of the reconstruction is provided by comparison with conventional strain scans from the KOWARI constant-wavelength strain diffractometer at the Australian Nuclear Science and Technology Organisation and simulations via finite element analysis.
\end{abstract}

\maketitle



\section{Introduction} 
\label{sec:introduction}
Bragg-edge neutron transmission techniques provide a means for obtaining lower dimensional (one- or two-dimensional) strain images from higher dimensional (two- or three-dimensional) strain fields within polycrystalline materials \red{\cite{santisteban02b,tremsin12}}.
The success of these techniques \red{\citep{santisteban2001time,priesmeyer1999bragg,tremsin2008energy,ramadhan2019characterization}}, and the development of associated instruments \red{\citep{johnson1997engin,shinohara2015commissioning,shinohara2016final,Tremsin_2010,carlile1999ess,Santisteban2002comparison,priesmeyer1994neutron}} and detectors \red{\citep{santisteban02b,tremsin2008energy,tremsin11}} has prompted research into the tomographic reconstruction of strain --- i.e. strain tomography.
This research aims to provide methods analogous to conventional Computed Tomography (CT) whereby the complete triaxial strain distribution within a sample could be reconstructed from a sufficient set of strain images. 
As the strain field is a tensor, this is a more complex task than conventional scalar CT.

If successfully developed, methods for the tomographic reconstruction of strain fields from these images could be used to study the residual elastic strain and stress within engineered components. 
Residual stresses are those which remain after applied loads are removed, for example due to heat treatment, or plastic deformation. Residual stresses may have a significant and unintended impact on a component's effective strength and service life, particularly its fatigue life.
A full field analysis of these stress and strains could have a significant impact on several areas of experimental mechanics.
In particular, it could be used to study the residual stress within additively manufactured, laser-clad, preened, welded, cast, forged and/or otherwise processed components. 
This full field analysis would have significant advantages over destructive and semi-destructive techniques (\citep{prime2001cross,standard2002standard,jang2003assessing,pang2003effects,leggatt1996development}), and \red{be complementary} to point-wise x-ray and neutron diffraction methods (\citep{hauk97,noyan87,fitzpatrick03,kisi2012applications,hutchings2005introduction}).

Strain tomography falls into the class of `rich' tomography problems where the \red{projected} strain image is related to an unknown tensor field.
The acquisition and analysis of these strain images is described in detail elsewhere \citep{santisteban02b,santisteban2001time,santisteban02}. Summarising this process: the relative transmission of neutron pulses with known wavelength-intensity spectra through the sample is measured at a pulsed neutron source (e.g. the Japan Proton Accelerator Research Complex (J-PARC) in Japan, ISIS in the United Kingdom, or the Spallation Neutron Source in the USA). 
For example, current state-of-the-art Micro-Channel Plate detectors \citep{tremsin11} are capable of measuring the transmitted spectra over an array of $512\times512$ pixels with a \red{pixel size} of $\SI{55}{\micro\metre}$. 
From this data, the position of a given Bragg-edge (a sudden increase in transmitted intensity as a function of wavelength) is observed at each pixel within the array.
The relative position of a Bragg-edge provides a measure of strain at each pixel of the form
\begin{equation}\label{eq:rel_strain_meas}
    \langle\epsilon\rangle = \frac{\lambda-\lambda_0}{\lambda_0},
\end{equation}
where $\lambda$ is the wavelength at which the Bragg-edge occurs, $\lambda_0$ is the corresponding Bragg-edge wavelength in a stress-free sample, and the following applies:
\begin{enumerate}
    \item As with all diffraction measurements, only the elastic component of strain is measured.
    \item The measured strain is the normal component in the direction of the transmitted neutron beam.
    \item The measurement corresponds to a through-thickness average along the path of the corresponding ray.
\end{enumerate}

\begin{figure}[htb]
    \centering
    \includegraphics[width=0.7\linewidth]{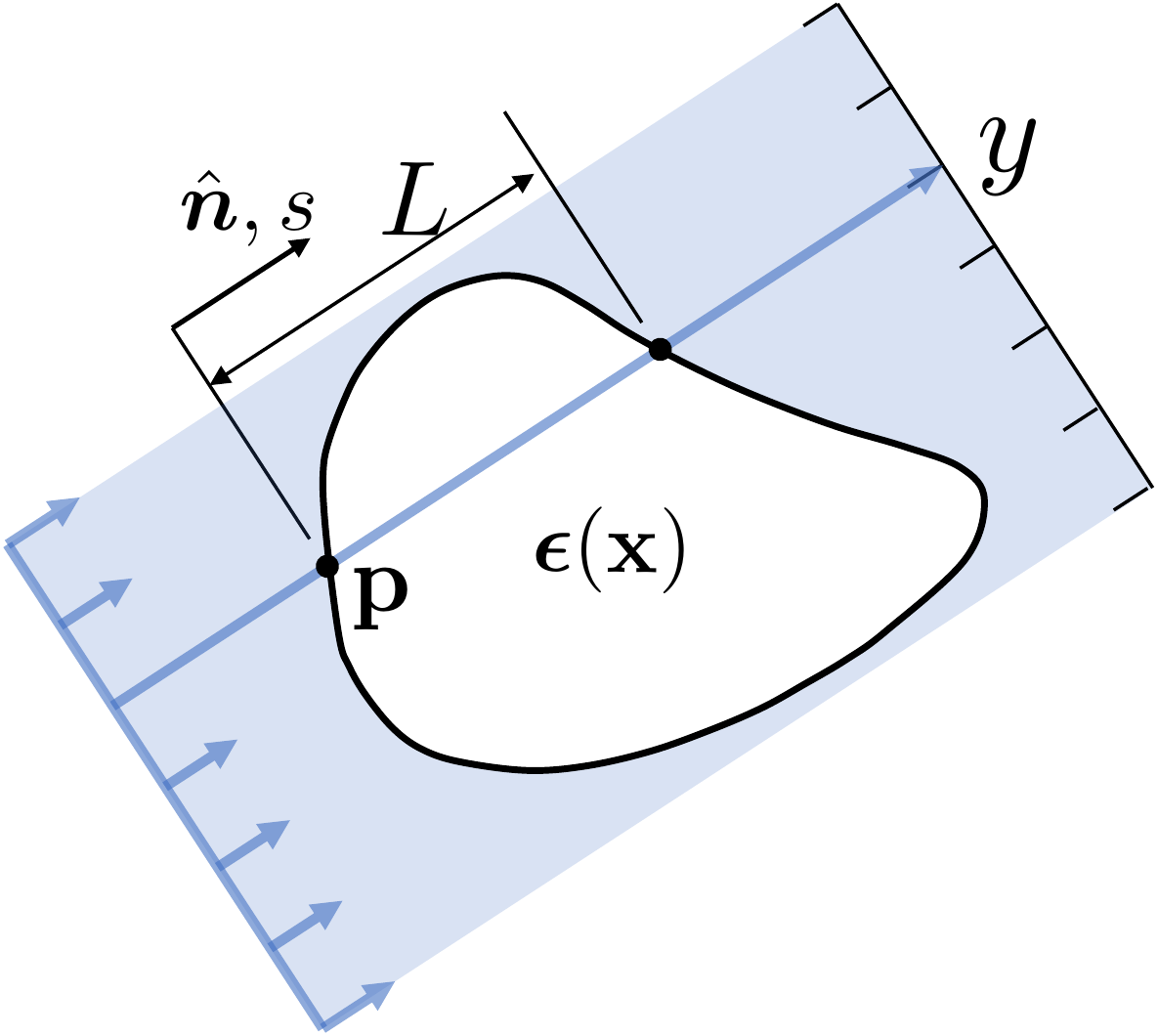}
    \caption{A two-dimensional representation of the Longitudinal Ray Transform. A ray of direction $\hat{\mathbf{n}}$ enters the sample at $\mathbf{p}$ and has a through-thickness length of $L$. The ray represents the path taken through the sample by neutrons arriving at the pixel with which the measurement is associated.}
    \label{fig:LRT}
\end{figure}

The strain measured at each pixel can be related to the strain field using the Longitudinal Ray Transform (LRT) \citep{abbey09,lionheart15}. With the inclusion of measurement error this gives a measurement model as
\begin{equation}\label{eq:LRT}
    y = \frac{1}{L}\int_{0}^{L}\hat{\mathbf{n}}^{\Transp} \boldsymbol\epsilon(\hat{\mathbf{n}}s+\mathbf{p})\hat{\mathbf{n}}\,\mathrm{d}s + e,
\end{equation}
where the LRT geometry is defined in Figure~\ref{fig:LRT} and $e$ is the measurement error term, which is assumed to be zero-mean Gaussian with standard deviation $\sigma$.
Estimating the strain field given a set of LRT measurements is made more complex as the LRT mapping is non-injective \citep{lionheart15,sharafutdinov1994integral}. 
This means that if the strain field components are considered independent, then infinitely many fields could produce the same set of measurements.
\red{The null space of the LRT poses a significant challenge to the development of methods for strain tomography.}

Strain tomography is an emerging field and research has been undertaken by a small number of groups, providing several methods to reconstruct two-dimensional strain fields. With the exception of \citep{kirkwood2019application} the majority of recent research has been undertaken by our own group.
Several special cases have been considered including axisymmetric systems \citep{abbey09,abbey12,kirkwood15,gregg2017tomographic,SATO2015349,kirkwood2019application} and granular systems \citep{wensrich16}.
More arbitrary strain fields caused by in-situ loadings have been successfully reconstructed by assuming compatibility \citep{wensrich2016bragg,hendriks2017bragg}. Methods for reconstructing a broader class of strain fields (i.e. residual strains caused by manufacturing processes) have been developed by ensuring the strain field satisfies equilibrium \citep{gregg2018tomographic,jidling2018probabilistic,hendriks2018traction}.
Outside the field of strain tomography, \red{neutron imaging has a range of applications from neutron radiography and tomography of geomaterials \citep{perfect2014hydrogen} to phase and texture imaging using diffraction contrast techniques \citep{woracek2018diffraction}, and \citet{anderson2009neutron} is a good reference for the interested reader.}

Recently, a \red{Gaussian process based approach suitable for modelling and estimating three-dimensional strain fields has been presented for simulated high-energy X-ray measurements \citep{hendriks2019bayesian}. 
Here, following the success of this approach, we present a Gaussian process based approach for neutron transmission strain tomography, and provide an experiment proof-of-concept demonstration.}



\section{Reconstruction Approach} 
\label{sec:approach}
The reconstruction approach is modified from the method presented by \citet{hendriks2019bayesian}.
This approach models the strain field by a Gaussian process (\citet{rasmussen2006gaussian} provides a good introduction), and ensures that the reconstructed strain field satisfies the physical properties of equilibrium; this method assumes the sample to be linearly elastic and isotropic (i.e. without texture).
Ensuring the strain field satisfies equilibrium is critical as the LRT mapping \eqref{eq:LRT} has a non-trivial null space \citep{lionheart15} (i.e. without these constraints a unique solution to the inverse problem does not exist).
By enforcing equilibrium the null space is reduced to contain only the trivial field, giving a unique solution to the problem \citep{hendriks2018traction}.

Gaussian processes are suitable for the modelling and estimation of spatially correlated phenomena.
The use of Gaussian processes to model and estimate strain fields was pioneered in \citet{jidling2018probabilistic}.
By modelling the Airy stress function, a scalar potential field, by a Gaussian process a solution to the two-dimensional stress (and hence strain) could be given that satisfies equilibrium in the absence of body forces.
This method can be extended to three dimensions for which the Beltrami stress functions are used instead of the Airy stress function. 
The Beltrami stress functions consist of six unique potential fields from which a complete solution to the equilibrium equations in three-dimensions can be given \citep{sadd2009elasticity}. 
These potential fields are each modelled by a Gaussian process allowing a tri-axial strain field satisfying equilibrium to be reconstructed.


In addition to equilibrium, boundary conditions can be included following the work by \citet{hendriks2018traction}.
Knowledge about unloaded surfaces for which the distribution of applied forces, known as tractions, is known to be zero can be incorporated.
This can be done by including artificial measurements of zero traction at points on the surface not subject to an applied load. This provides information about the stress of the form
\begin{equation}
    \mathbf{0} = \mathbf{n}_\perp^{\Transp} \boldsymbol\sigma(\mathbf{x}_b),
\end{equation}
where $\mathbf{x_b}$ is a point on an unloaded surface, $\boldsymbol\sigma$ is the triaxial stress field, and $\mathbf{n}_\perp$ is the unit vector perpendicular to the surface. Hooke's law can then be used to relate this information to the strain field, improving the reconstruction near the samples' boundary.

A detailed description of the method and its implementation is given by \citet{hendriks2019bayesian}. This requires only minor modification to the measurement model, and the inclusion of traction measurements.
The measurement model requires a slight modification due to the difference between high-energy x-ray and neutron transmission strain measurements.
For high-energy x-ray strain measurements, the measured strain direction, denoted $\hat{\boldsymbol{\kappa}}$, is almost perpendicular to the ray direction $\hat{\mathbf{n}}$.
Whereas, for neutron transmission strain measurements the direction of measured strain is aligned with the ray, and so $\hat{\boldsymbol{\kappa}} = \hat{\mathbf{n}}$.
Additionally, a large variation in strain measurement uncertainty is observed and therefore the method is modified so that each measurement can be assigned its own standard deviation. 
In essence this weights the importance of each measurement according to its confidence given by the strain imaging process.
Details of these modifications are given in in Appendix~\ref{sec:modifications_to_the_method}.

In this paper, this approach is used to reconstruct the strain field within an EN26 steel sample from a set of strain images. The sample, load case, and strain image acquisition are described in Section~\ref{sub:sample_design_and_data_acquisiation}. 
The resulting strain field is validated by comparison with conventional diffraction strain scans and FEA results, which are described in Section~\ref{sub:validation}.
The reconstructed strain field and a comparison with the validation data is given in Section~\ref{sub:results}; potential sources of error are also discussed in this section.

\label{sec:demonstration_experimental}
\subsection{Sample Design and Strain Imaging} 
\label{sub:sample_design_and_data_acquisiation}
\begin{figure*}[htb]
    \centering
    \includegraphics[width=0.225\linewidth]{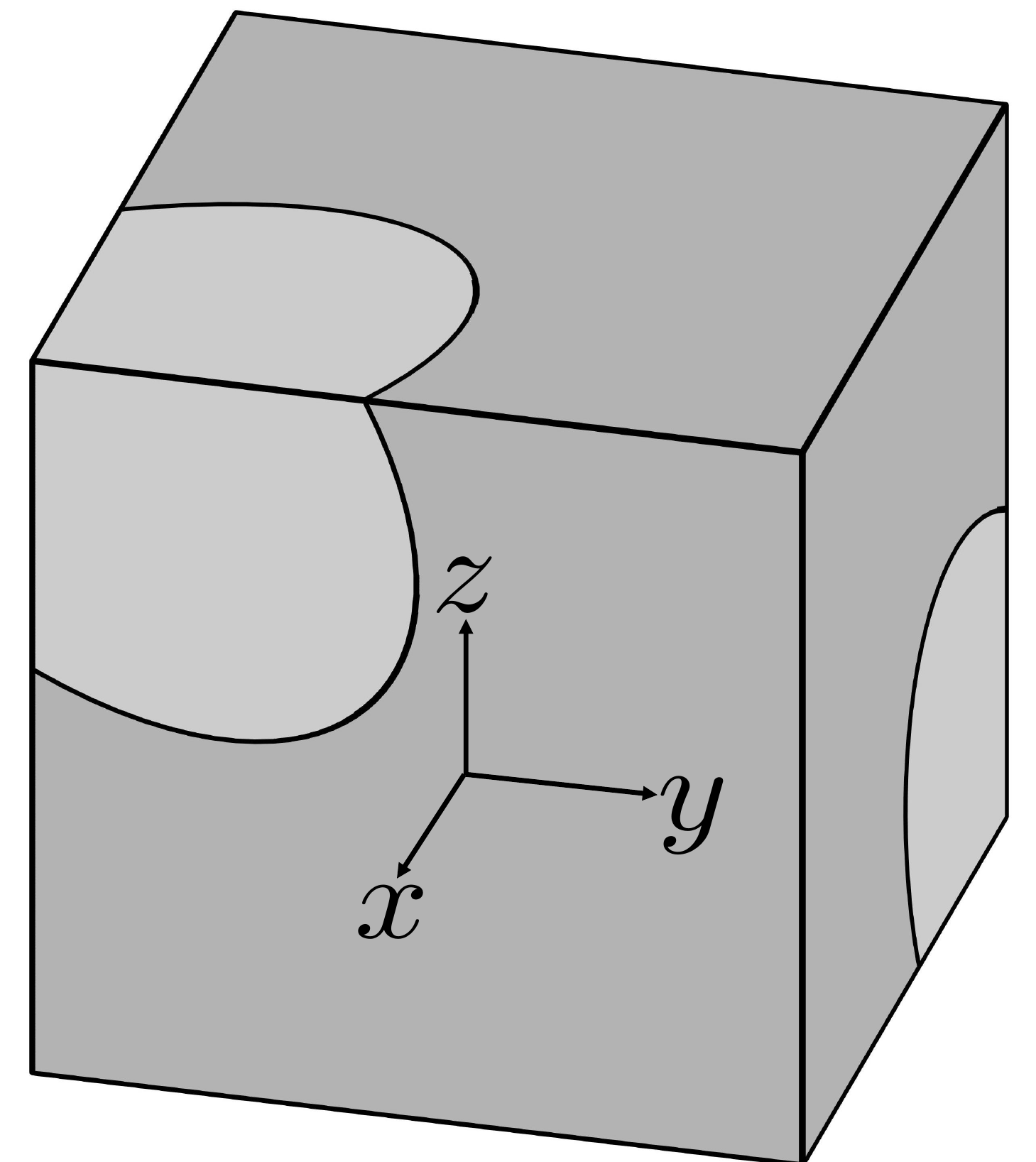}
    \includegraphics[width=0.225\linewidth]{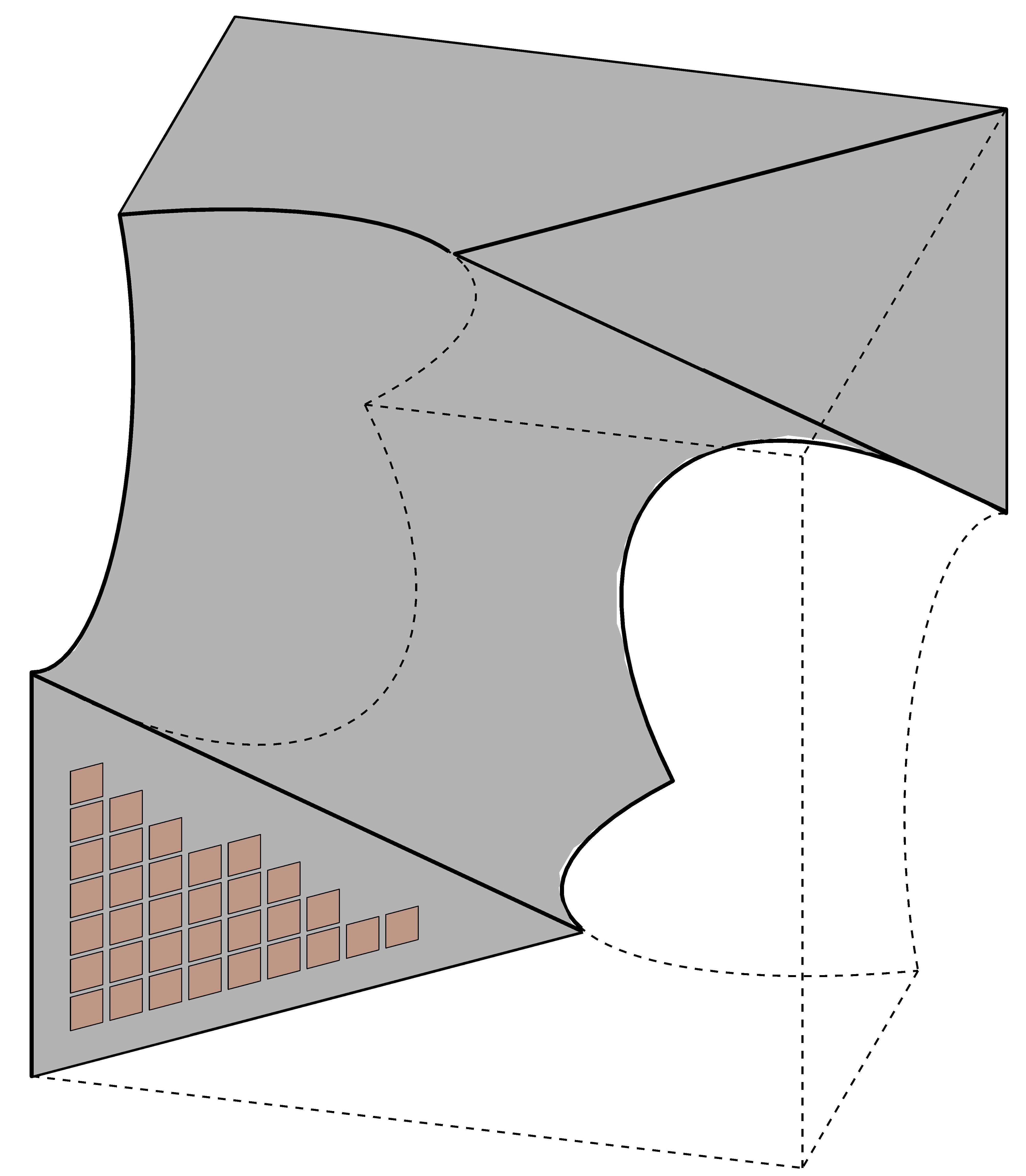}
    \includegraphics[width=0.225\linewidth]{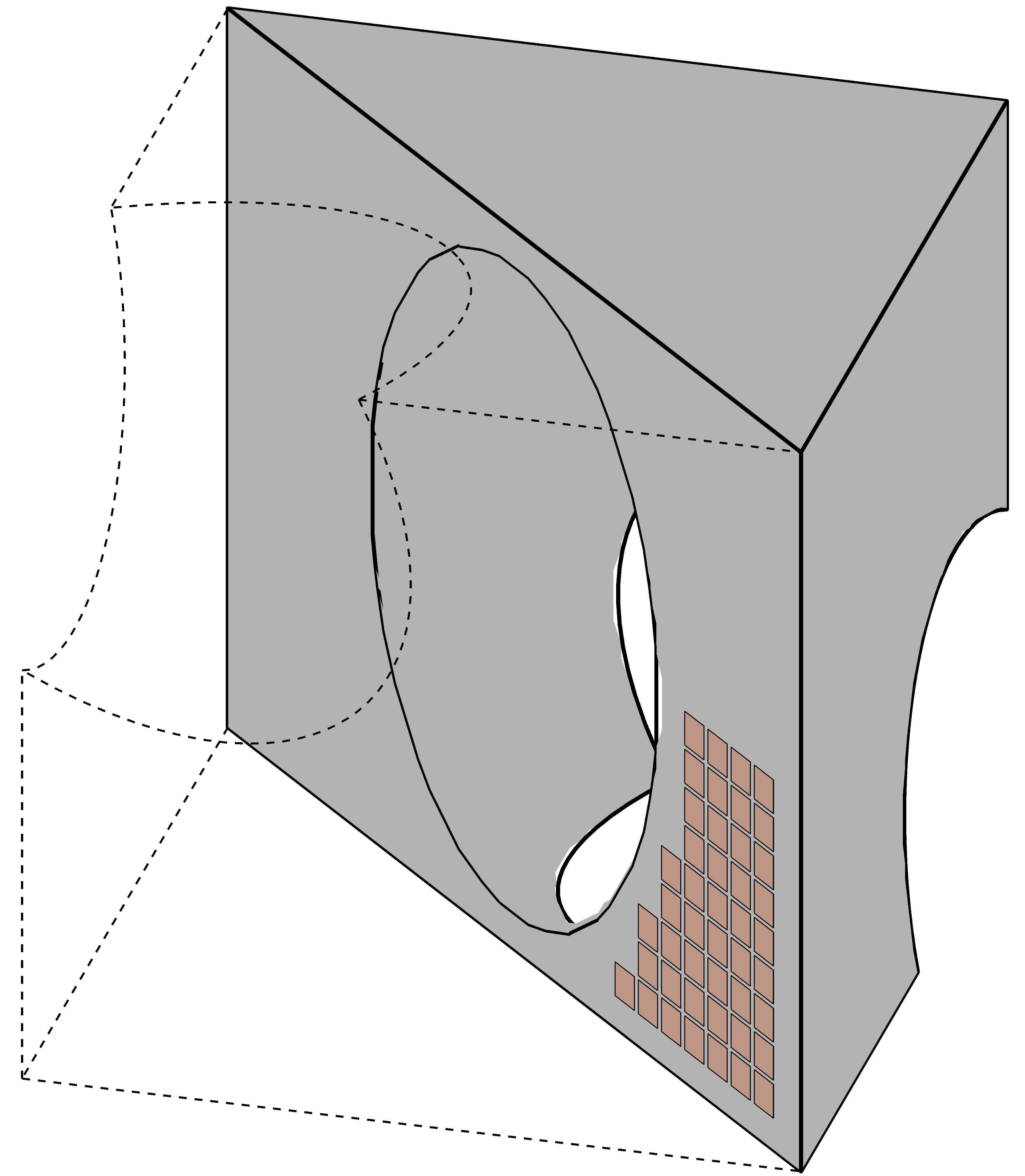}
    \includegraphics[width=0.225\linewidth]{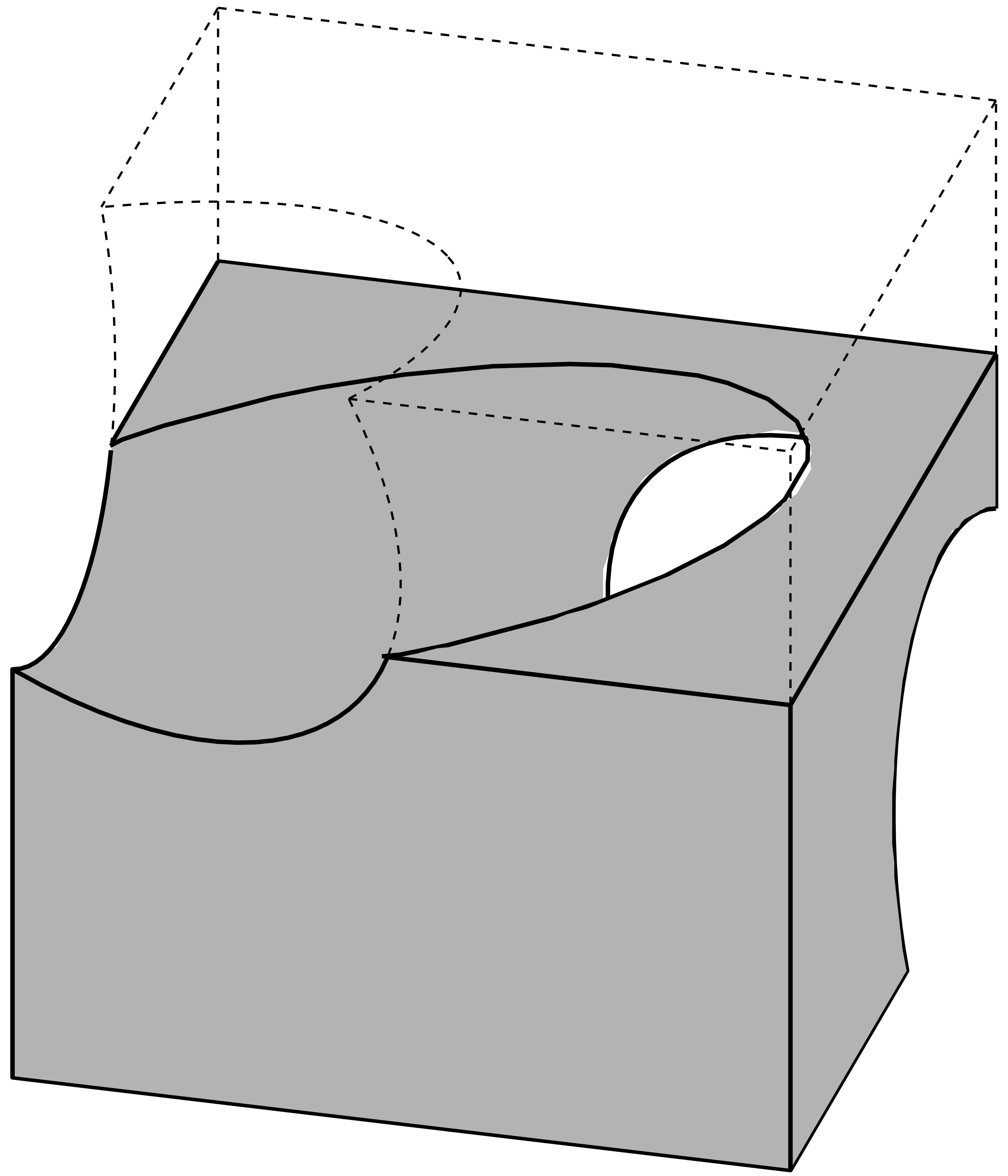}
    \caption{Left: The sample (dark grey) and plug (light grey) assembly, as well as the coordinate systems used. Middle Left: Cross section plane 1 used for validation with KOWARI and FEA. Middle Right: Cross section plane 2 used for validation against KOWARI and FEA. Right: Cross section plane plane 3 used for validation against FEA. For section planes one and two, the location of the KOWARI measurements are shown in orange. Note that gauge volume orientation shown is indicatively only, as it varies for each component of strain measured.}
    \label{fig:sample}
\end{figure*}

The method is applied to a set of strain images collected on the RADEN energy-resolved-neutron-imaging instrument at J-PARC \citep{shinohara2015commissioning,shinohara2016final} of an EN26 steel (medium carbon, low alloy) sample. 
The sample consisted of a $17\times17\times\SI{17}{\milli\meter}$ steel cube with a precision ground hole of diameter $\SI{12}{\milli\meter}$ along the diagonal. A load was applied by a $40\pm\SI{2}{\micro\meter}$ interference fit, i.e. shrink fit, with a titanium plug. 
The sample and plug are shown in Figure~\ref{fig:sample}.

Prior to assembling, the sample was heat treated to relieve stress and provide a uniform tempered-martensite (i.e. ferritic) structure with minimal texture, and with a final hardness of $\SI{290}{HV}$.
The sample was assembled by first inserting the plug into a cylinder with an interference fit of $40\pm\SI{2}{\micro\metre}$, after which a cube with $\SI{17}{\milli\metre}$ sides was milled from the cylinder and plug.

This sample and loading set-up was designed to provide a smooth three-dimensional strain field suitable for the first demonstration of three-dimensional strain tomography. 
This has additional advantages for validation as FEA of strain resulting from interference fits is a more straight forward process than FEA of strain fields resulting from plastic deformation.

\begin{figure}[htb]
    \centering
    \includegraphics[height=6cm]{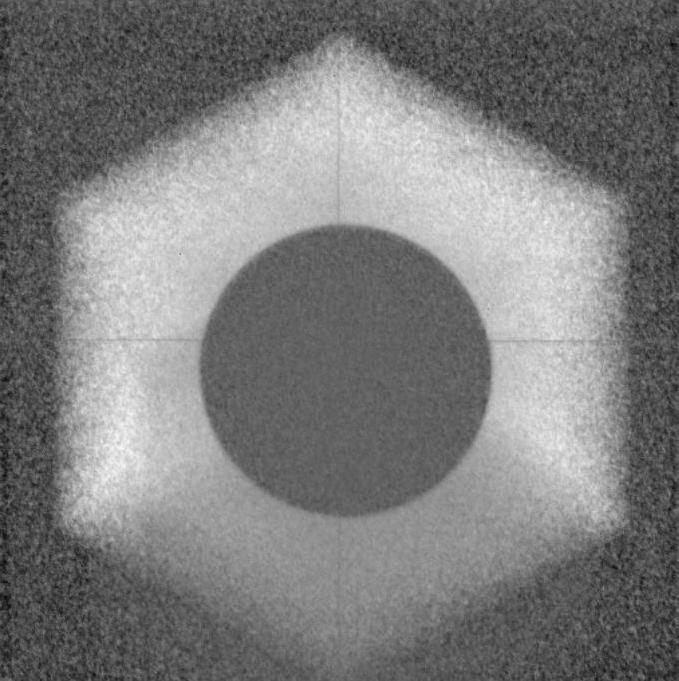}
    \caption{Bragg-edge height map for a projection that aligned the beam direction and the plug. This illustrates that the titanium plug does not contribute to the Bragg-edge chosen for analysis.}
    \label{fig:edge_height}
\end{figure}

To this end, a titanium plug was chosen as strain within the plug was `invisible' to the strain imaging process.
This is because titanium does not have a Bragg-edge near to the steel Bragg-edge that was chosen for analysis (see Figure~\ref{fig:edge_height}).
As a result, the titanium plug can be ignored during the reconstruction process and considered as imparting an in-situ load on the interior face of the hollow cube.

Strain images were measured using a micro-channel plate detector ($512\times512$ pixels, $\SI{55}{\micro\metre}$ per pixel)\citep{tremsin11} at a distance of $\SI{17.9}{\metre}$ from the source. Each image required $2.5$ hours of beamtime. At the time of the experiment (January 2019) the source power was $\SI{500}{\kilo\watt}$.
Pixels were grouped together into macro pixels of $24\times24$ giving sufficient neutron counts to provide a reasonable edge fit; giving a final strain image of $21\times21$ macro pixels, each with a resolution of $\SI{1.3}{\milli\metre}$.
It is worth noting that this does correspond to the resolution of the final reconstruction and this is discussed more in Section~\ref{sub:results}.


Each macro pixel provides a strain measurement of the form \eqref{eq:rel_strain_meas} where the Bragg-edge position was found following the procedure given by \citet{santisteban2001time} applied to the (110) Bragg-edge. The undeformed location $\lambda_0$ was determined from a stress-free sample.
The resulting strain measurements had, on average, an uncertainty of standard deviation $\sigma=\num{2.7d-4}$.
This measurement uncertainty is higher than previously achieved in two-dimensional strain tomography experiments \citep{hendriks2017bragg,gregg2018tomographic}. However, further increasing the macro pixel size did not provide a sufficient decrease in uncertainty to warrant the loss in strain image resolution.
Additionally, a systematic bias in the edge fit as a function of the measurement path length was observed. 
This effect was previously described in \citet{vogel2000tof} and \citet{gregg2018tomographic}; although the exact mechanism is yet to be established.
Following \citet{gregg2018tomographic}, a linear correction was applied to $\lambda_0$.

\begin{figure*}[htb]
    \centering
    \subcaptionbox{Sample positioning \label{fig:sample_positioning}}{
    \includegraphics[height=6cm]{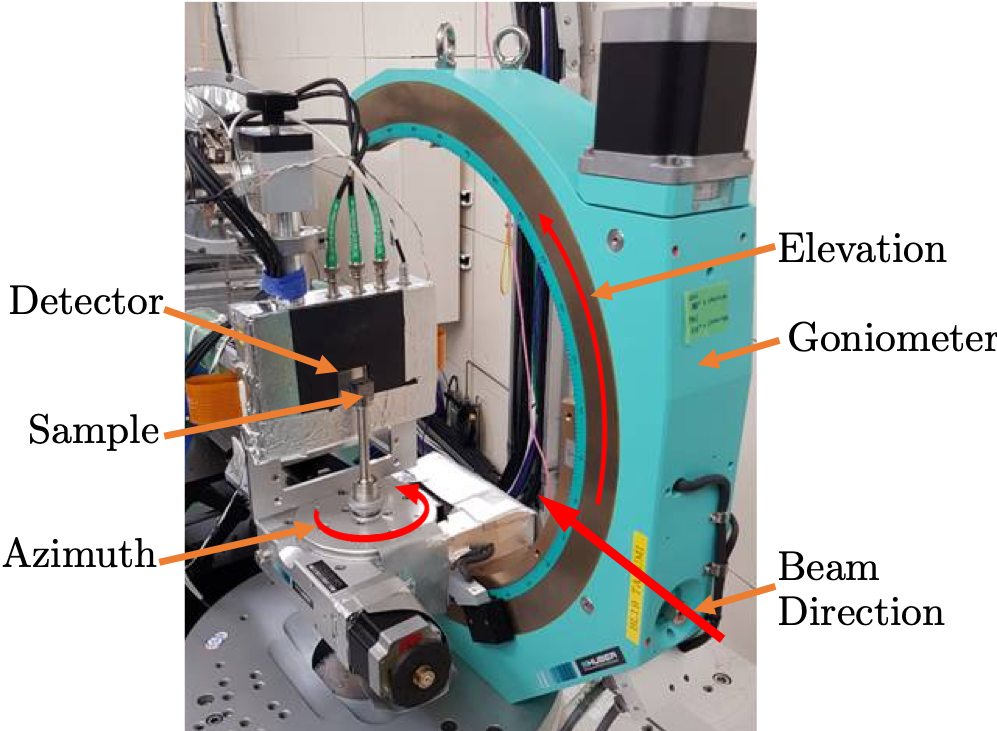}} 
    \subcaptionbox{Measurement directions \label{fig:proj_directions}}{\includegraphics[height=6cm]{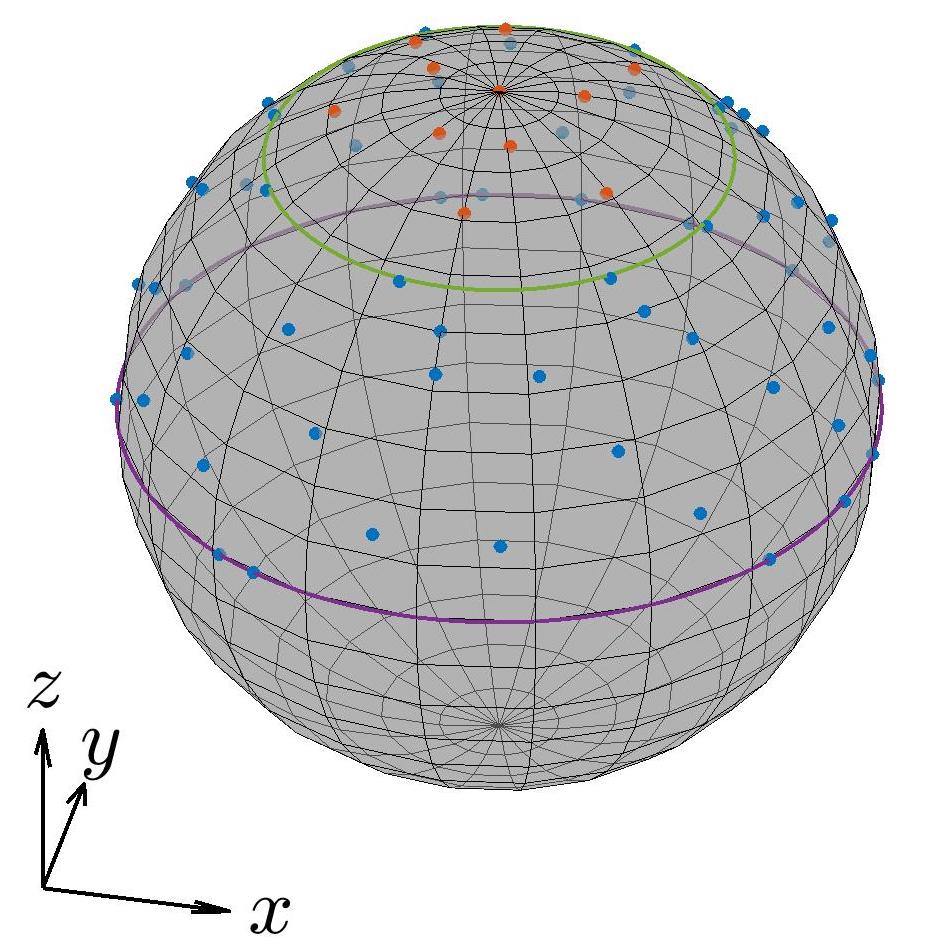}}
    \caption{(\subref{fig:sample_positioning}) Sample, detector, goniometer, and beam direction are shown. The geniometer enabled rotation in azimuth and elevation. The $x$-axis of the sample was aligned with the beam and the $z$-axis aligned with vertical. (\subref{fig:proj_directions}) The measurement directions corresponding to the azimuth and elevation used for the strain images. 
    The coordinates of the blue and orange points correspond to the $x,y,z$ components of the direction unit vectors. The region between the purple and green line represents the achievable elevation range of 0 to $52^\circ$. As such, measurement directions represented by the orange points required the sample to be rotated.}
    \label{fig:experiment_setup}
\end{figure*}

In total 70 strain images were collected. For these images the sample was positioned using a two axis goniometer, which allowed rotation in azimuth and elevation (see Figure~\ref{fig:experiment_setup}\subref{fig:sample_positioning}). 
Limitations of the experimental set-up restricted the achievable elevation angles to the range from $0$ to $52^\circ$.
Therefore, in order to cover the full range of measurement directions, the sample was repositioned by rotating $90^\circ$ about the $y$-axis for the final 11 strain images.
The measurement directions corresponding to azimuth and elevation angles used are shown in Figure~\ref{fig:experiment_setup}\subref{fig:proj_directions}.
Since the LRT is symmetric, measurements with opposite directions provide the same information. Hence, only measurement directions covering one hemisphere are required.

The exact geometry of each neutron ray passing through the sample before reaching the detector is required to model the measurements by the LRT \eqref{eq:LRT}.
In addition to designing a sample holder to carefully position the sample, an optimisation routine was run to determine the remaining orientation offsets  and the offsets between the centre of rotations.
The optimisation maximised the sum of Bragg-edge heights associated with rays that would pass through the sample for a given choice of offsets.

Strain fields were reconstructed from this set of strain images using the Gaussian process method described earlier with the inclusion of 400 measurements of zero traction evenly distributed on each of the exterior faces. 


\subsection{Validation Data} 
\label{sub:validation}

Validation relies on comparison with conventional strain scans \citep{kisi2012applications,fitzpatrick03,noyan87} from the KOWARI constant-wavelength strain diffractometer at the Australian Nuclear Science and Technology Organisation \citep{kirstein2009strain,brule2006residual} and Finite Element Analysis (FEA).
The strain scans provide measurements of the 6 components of strain on two section planes (33 points on plane 1 and 45 points on plane 2).
As with all diffraction methods, these measurements correspond to the average strain inside gauge volumes.
These gauge volume locations were chosen on two section planes that were expected to exhibit strong skew symmetry and therefore help to validate a larger region of the reconstruction. 
The section planes and gauge volume locations are shown in Figure~\ref{fig:sample}.

These measurements were based on the relative shift of the (211) diffraction peak measured with neutrons of wavelength $\lambda = \SI{1.67}{\angstrom}$ ($\ang{90}$ geometry) and a $1.0\times1.0\times\SI{1.0}{\milli\meter^3}$ gauge volume. The $\{211\}$ and $\{110\}$ lattice planes have effectively the same diffraction elastic constants \citep{daymond2002elastoplastic}, therefore the results from the transmission and diffraction experiments can be directly compared without rescaling or recalculation to stress.
Sampling times with the KOWARI diffractometer were based on providing uncertainty in strain around $\num{1e-4}$: which required 60 hours.
The long sampling times required for the comparatively small gauge volumes meant that only a portion of each section plane could be measured.

While only a small amount of the strain field can be verified using the strain scanning measurements, the full reconstructed strain field can be compared to FEA results to provide further validation.
Comparison of the reconstruction to the FEA results is made for the three section planes shown in Figure~\ref{fig:sample}.


\subsection{Results and Discussion} 
\label{sub:results}
\begin{figure*}[htb]
    \centering
     \includegraphics[width=0.8\linewidth]{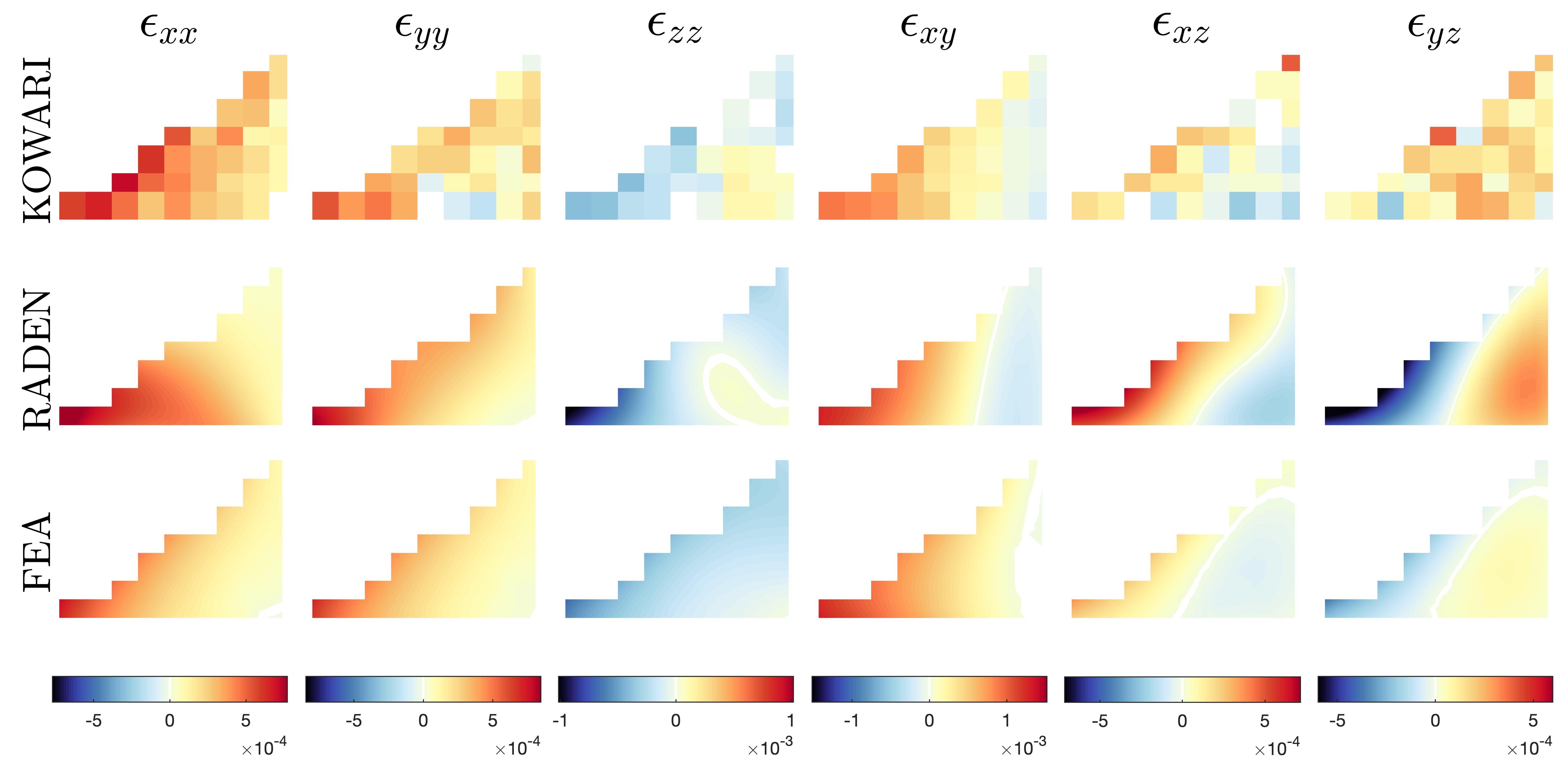}
    \caption{Comparison of the KOWARI strain scans, the reconstruction from RADEN strain images, and FEA results. Shown is the region on section plane 1 for which KOWARI strain scans were made. The KOWARI strain scans correspond to measured averages within the $1\times1\times\SI{1}{\milli\meter^3}$ gauge volumes and so are shown as a constant value within representative gauges to best reflect this.}
    \label{fig:plane1_KOWARI_comp}
\end{figure*}

\begin{figure*}[htb]
    \centering
    \includegraphics[width=0.8\linewidth]{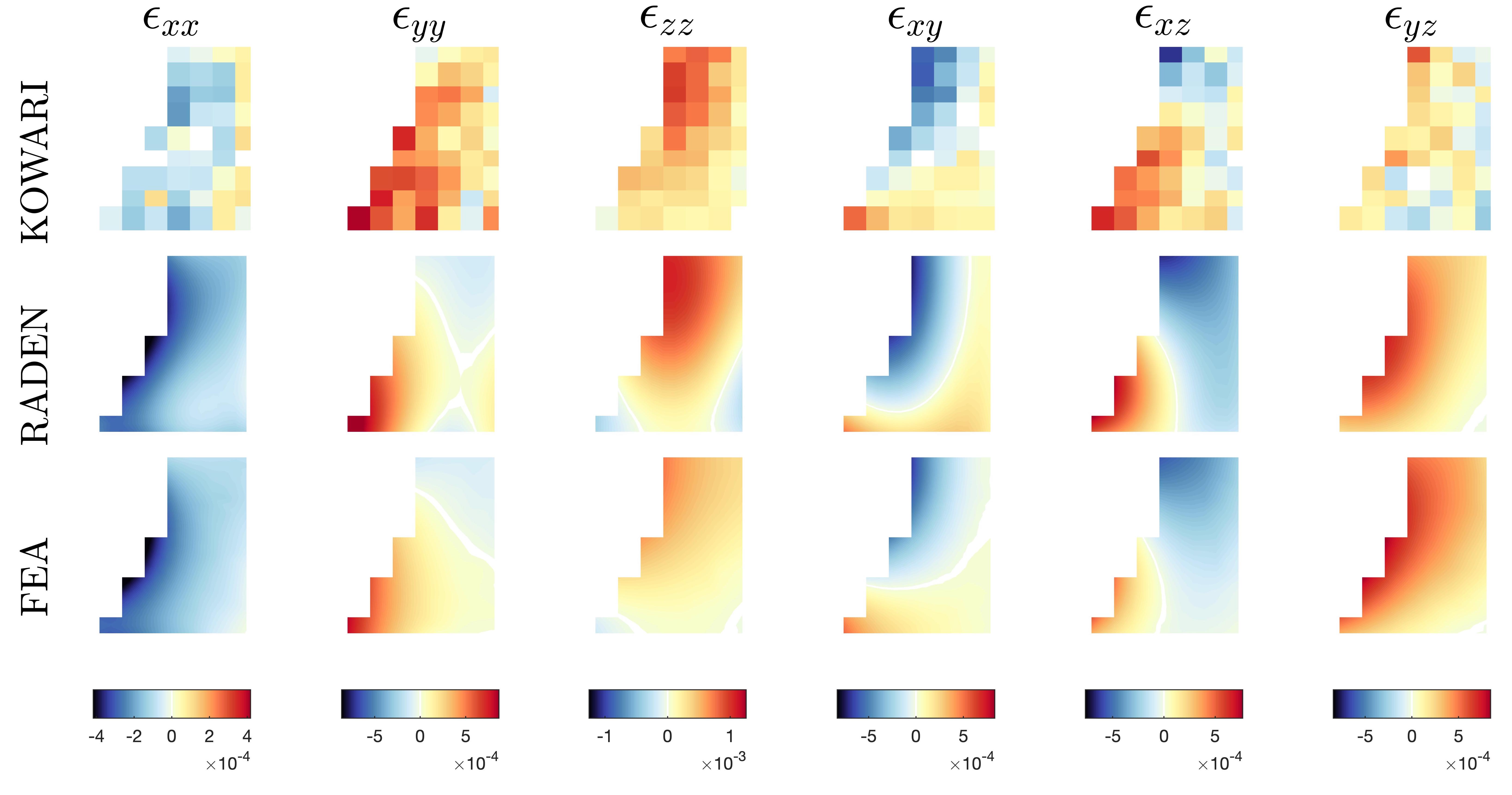}
    \caption{Comparison of the KOWARI strain scans, the reconstruction from RADEN strain images, and FEA results. Shown is the region on section plane 2 for which KOWARI strain scans were made. The KOWARI strain scans correspond to measured averages within the $1\times1\times\SI{1}{\milli\meter^3}$ gauge volumes and so are shown as a constant value within representative gauges to best reflect this.}
    \label{fig:plane2_KOWARI_comp}
\end{figure*}

The reconstructed strain field is shown together with strain scans made on the KOWARI diffractometer and FEA results in Figures~\ref{fig:plane1_KOWARI_comp}~and~\ref{fig:plane2_KOWARI_comp}.
As the KOWARI strain scans cover only a region of section plane 1 and 2, comparison between all three sets is made in these regions only.
Example Matlab code to perform this reconstruction and produce the strain plots shown in this paper is available on Github \citep{githubcode}.

Comparison of the KOWARI strain scans, the RADEN reconstruction and the FEA results shows good agreement in general. 
However, some specific differences can be noted:
\begin{itemize}
    \item In the $\epsilon_{zz}$ component of plane 1, the reconstruction shows a region of tension in the bottom right that is also present in the strain scans but is not seen in the FEA.
    \item In the $\epsilon_{yz}$ and $\epsilon_{xz}$ components of plane 1, the reconstructed strain has the same shape but a greater magnitude than the FEA strain and it is not clear whether this is supported in the strain scans.
    \item In the $\epsilon_{yy}$ component of plane 2, the reconstructed strain field has a region of compression also present in the FEA but not seen in the strain scans.
\end{itemize}

When making these comparison it is important to remember that the KOWARI strain scans are themselves measurements which are relatively noisy ($\num{1e-4}$ standard deviation) and are averages over the gauge volumes; this in some cases makes comparison difficult.
In particular, the $\epsilon_{xz}$ and $\epsilon_{yz}$ components of plane 1, and the $\epsilon_{xx}$ and $\epsilon_{yz}$ components of plane 2 appear particularly noisy.

\begin{figure*}[htb]
    \centering
    \includegraphics[width=0.6\linewidth]{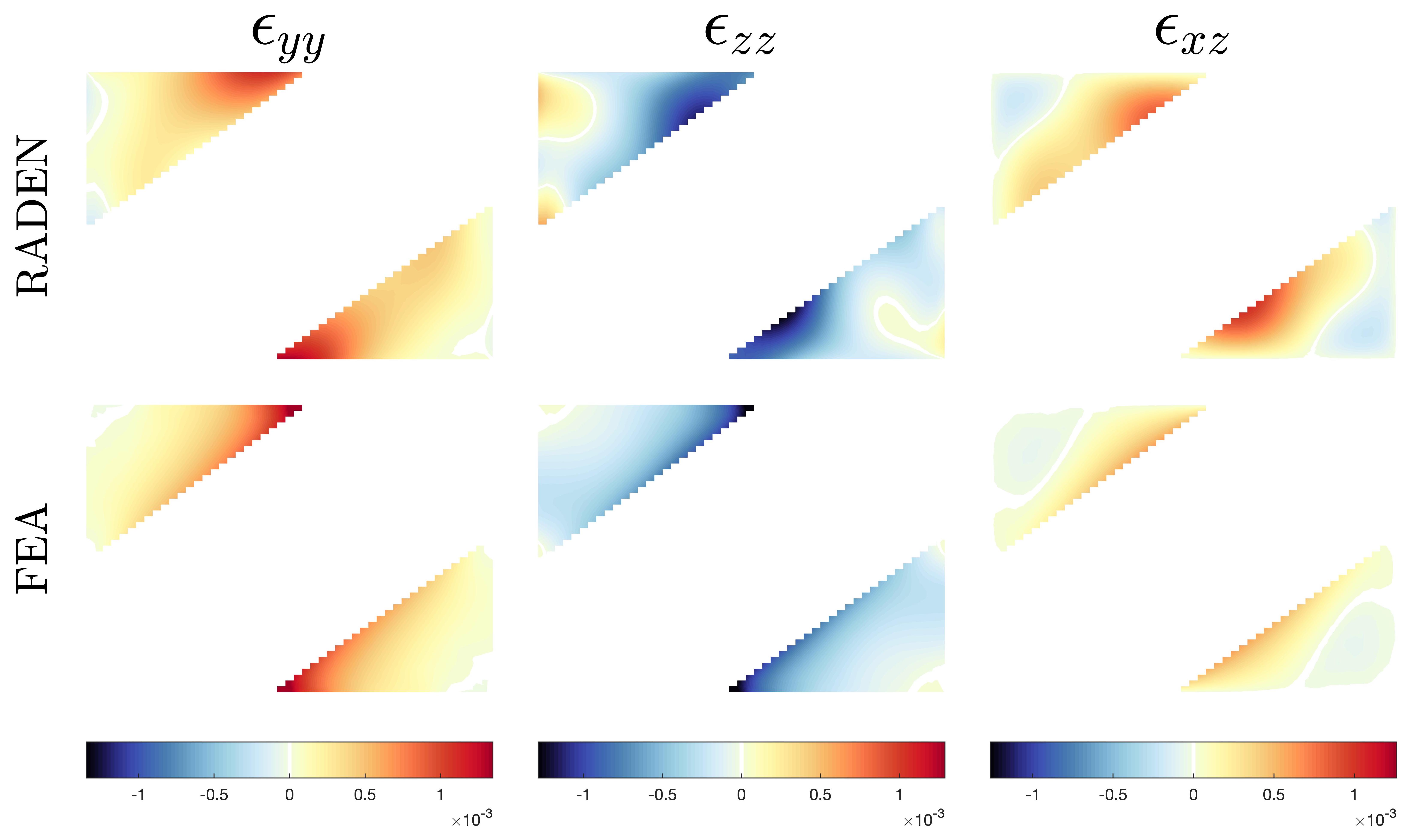}
    \caption{The strain field reconstructed from the RADEN strain images and the FEA strain field for section plane 1.}
    \label{fig:plane1_complete}
\end{figure*}

\begin{figure*}[htb]
    \centering
    \includegraphics[width=0.6\linewidth]{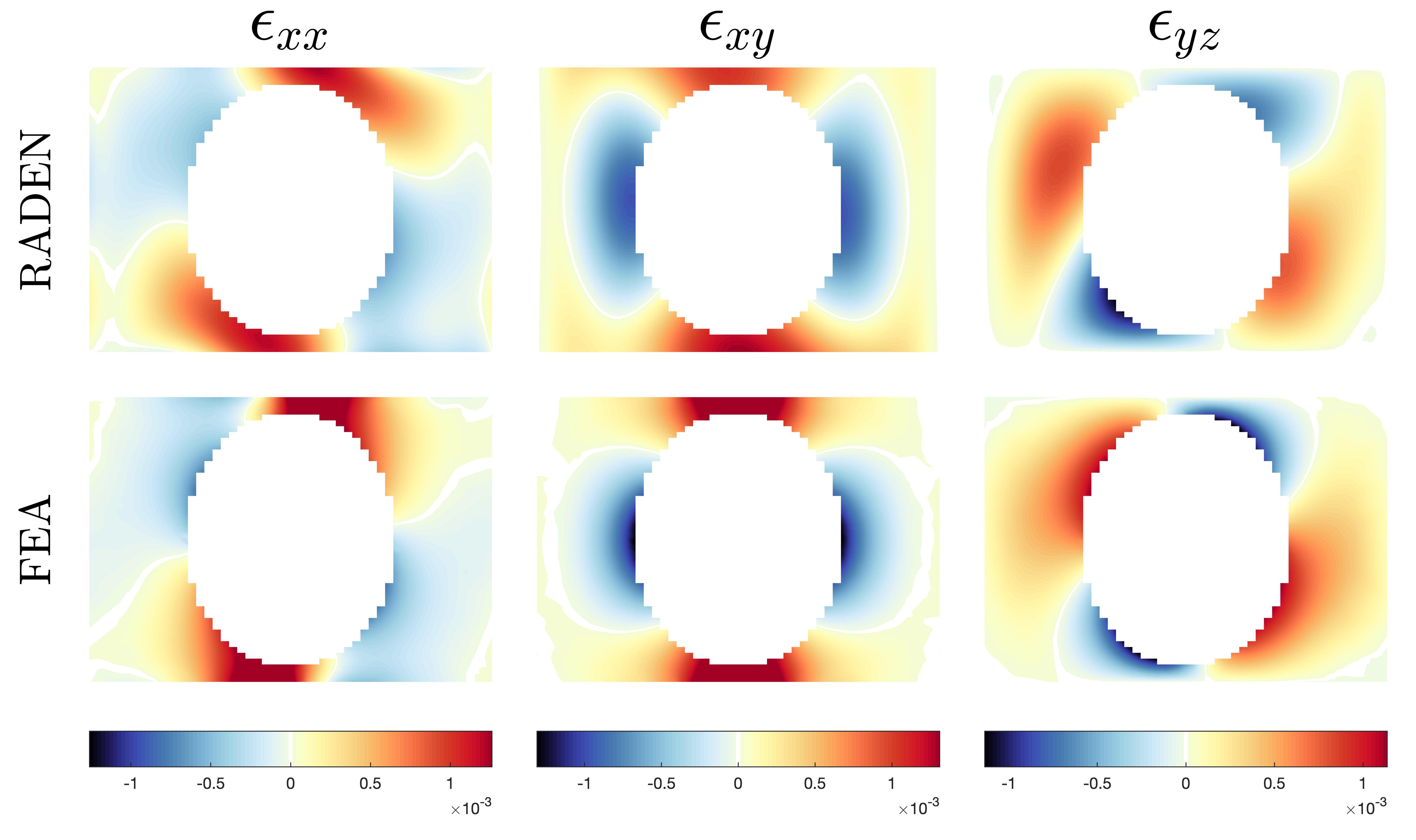}
    \caption{The strain field reconstructed from the RADEN strain images and the FEA strain field for section plane 2.}
    \label{fig:plane2_complete}
\end{figure*}

\begin{figure*}[htb]
    \centering
    \includegraphics[width=0.6\linewidth]{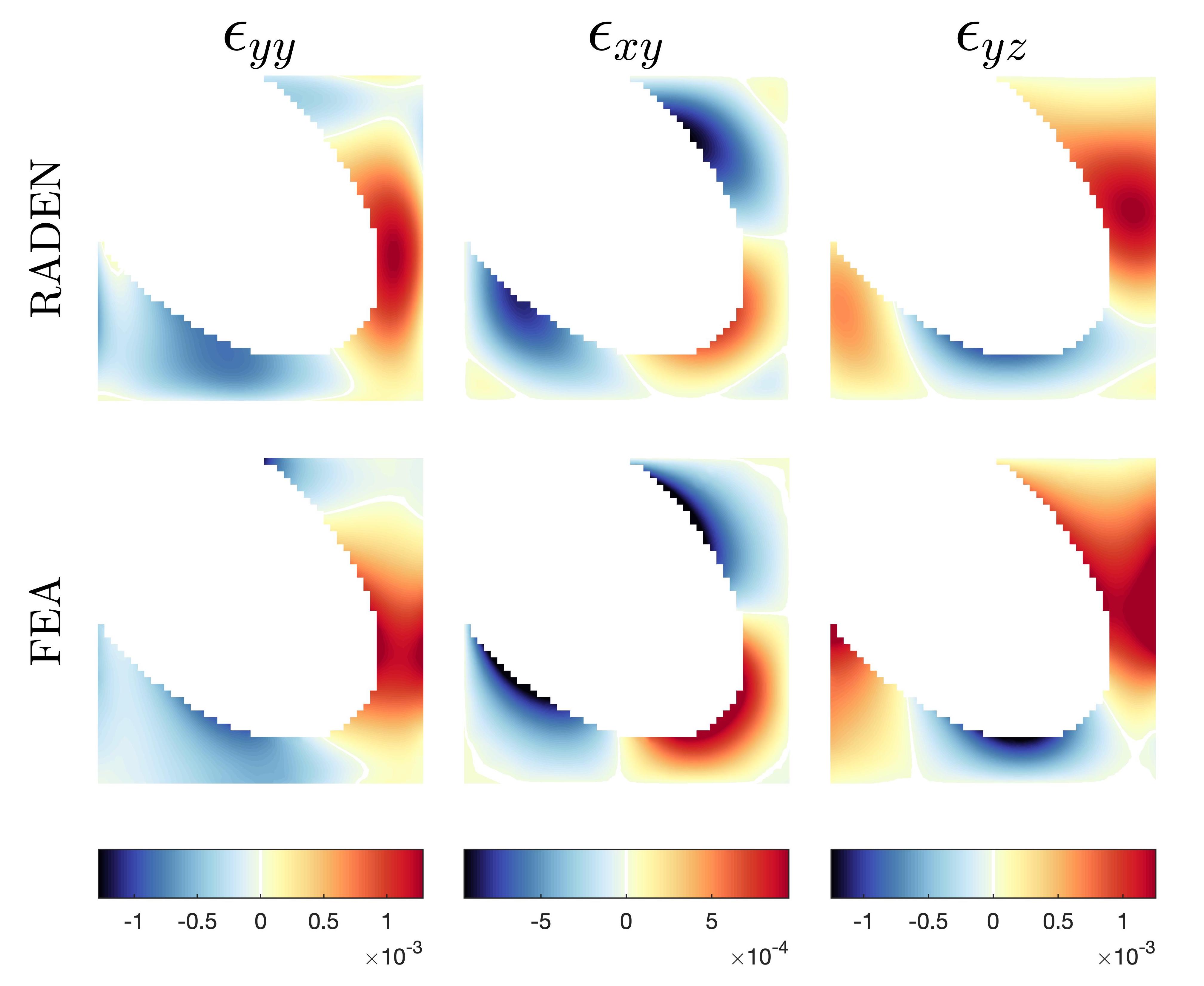}
    \caption{The strain field reconstructed from the RADEN strain images and the FEA strain field for section plane 3.}
    \label{fig:plane3_complete}
\end{figure*}

The FEA strain fields and the reconstructed strain fields are available for the entire sample allowing comparison over the entire section planes 1, 2, and 3; these are shown in Figure~\ref{fig:plane1_complete}, Figure~\ref{fig:plane2_complete} and Figure~\ref{fig:plane3_complete} respectively.
A selection of strain components is shown for each section plane so that all components are shown at least once. 
This comparison indicates that the reconstruction shows close agreement with the FEA strain fields on plane 2 and plane 3. 
The shape of the reconstructed strain field is very similar with the main difference being noted in slightly reduced magnitudes and the peak strains are less concentrated. 
There is a slightly greater difference between the reconstruction and FEA results observable in plane 1, particularly in the $\epsilon_{xz}$ and $\epsilon_{zz}$ components. 
However, at least for the $\epsilon_{zz}$ component some of this difference is supported by the KOWARI strain scans which also showed the region of tension present in the bottom right. 
Additionally, as the strain fields are skew symmetric this would lend some support to the region of tension present in the top left of this component.

From these results, a quantitative assessment of the discrepancies between the the KOWARI strain scan measurements and the reconstruction as well as between the reconstruction and the FEA results was carried out. 
In both cases the differences are mean zero and Gaussian, implying that there is no systematic error or bias resulting from the reconstruction technique.
The differences between the KOWARI strain scans and the reconstruction are calculated by taking gauge volume style averages of the reconstructed strain field and comparing these to the KOWARI measurements. 
The resulting differences have an average magnitude of $\SI{48}{\micro\text{Strain}}$. 
The discrepancies between the FEA and reconstructed strain fields have an average magnitude of $\SI{145}{\micro\text{Strain}}$. 

Although the reconstruction shows, in general, good agreement with the FEA results and KOWARI strain scans some differences have been noted and these differences may be attributed to several sources.
Firstly, the number and quality of strain measurements acquired is less than has been previously achieved in two-dimensional strain tomography experiments. 
In total, 14000 strain measurements were acquired with an average standard deviation of $\num{2.7e-4}$ compared to approximately $20000$ with a standard deviation of $\num{1e-4}$ in previous two-dimensional experiments \citep{hendriks2017bragg,gregg2018tomographic}.

This is in part due to the trade off between macro pixel size and uncertainty in the Bragg-edge fits. While smaller macro pixels ($16\times16$ pixels) would give better resolution in the strain image, the measurement standard deviations would be increased to around $\num{3.4e-4}$. 
Conversely, larger macro pixels ($32\times32$) could be used to decrease the measurement standard deviation to around $\num{2.3d-4}$, however the resolution in the strain image would be made worse.
This is in contrast to two-dimensional geometry where the assumption of no out of plane strain variation meant the pixels could be binned into columns without affecting the resolution of the resulting one-dimensional strain image.

It is important to highlight that this measurement uncertainty does not directly correspond to the uncertainty in the reconstructed strain field. 
Since the reconstruction relies on combining information from multiple strain images, it can have lower uncertainty than the individual strain measurements.
A benefit of the Gaussian process method used is that it also gives an estimate of reconstruction's standard deviation (i.e. the uncertainty in the estimate).
For the presented reconstruction, the average standard deviation was $\num{4.1d-5}$, which is marginally better than the neutron diffraction strain measurements used for validation.

Further, the resolution of the strain images, given by the macro pixel size, does not explicitly correspond to the resolution of the reconstruction. Although analogous to conventional CT, the reconstruction method used is fundamentally different. This method does not break the region into discrete voxels, instead it provides a continuous smooth estimate of the strain field where the maximum rate of change is automatically adapted to suit the available data. This means the idea of a reconstruction resolution does not directly apply. Instead, we could look at the maximum strain gradient that can be captured in the reconstruction. For instance, by applying the method to a sample with a step change in strain and analysing the distance over which this change occurs in the reconstruction an equivalent `resolution' could be determined. Since the LRT is only defined within the sample, this step change must be internal, ruling out the application of this idea to our sample. 
\red{While the reduced peak magnitudes are in part due to the inherent smoothing and as such related to an equivalent resolution, these reductions in magnitude are also due to low certainty in measurements of these regions as a result of small Bragg-edge heights, and macro pixel binning. Hence, an important study for future research would be to apply the method to a sample with an internal step change in strain.}

The impact of the binning to form macro pixels is particularly significant in regions near the sample boundary, for example in the corners where the plug and sample intersect. These regions contain a small amount of material and so the averaging effect of the LRT means that the strains are poorly sampled. Additionally, the smaller amount of material means that the Bragg-edge height of any strain measurements passing predominately through these regions is reduced, resulting in poorer measurement confidence. When combined with the macro pixel averaging \red{and inherent smoothing of Gaussian process method}, this results in the peak strains in these regions being somewhat obscured.
This issue may be somewhat alleviated in future experiments at J-PARC with an expected source power increase to $\SI{1}{\mega\watt}$ over the next few years.

Differences between reality and the FEA model may also account for some of the observed discrepancies. The sample was milled into a cube from a typical ring and plug. During this process it was not possible to ensure the plug was perfectly on the diagonal of the cube.
This would account for the measured strain fields (KOWARI and RADEN) not being entirely skew symmetric and for some of the difference between these fields and the FEA strain field.
Additionally, the milling process itself may have introduced residual stresses not accounted for in the FEA model.
Finally, the peak stresses are around the yield strength of the material which could result in effects such as hardening and account for differences between the measured strain fields and the FEA strain field.


\section{Conclusion and Future Work} 
\label{sec:conclusion}
A proof-of-concept demonstration for triaxial strain reconstruction from neutron transmission strain images has been provided.
Strain images were collected using the RADEN energy-resolved-neutron-imaging instrument at J-PARC.
The reconstructed strain field was validated by comparison with conventional strain scans from the KOWARI diffractometer and FEA results and shows good agreement.

The reconstruction was performed using a Gaussian process based method that ensures the resulting strain field satisfies equilibrium. 
This is achieved by using the Beltrami stress functions to provide a complete solution to the stress (and strain) fields in three-dimensions. 
The reconstruction provides a smooth, continuous estimate of the strain field throughout the entire sample.

The reconstructed strain field was developed within a hollowed EN26 cube by an `in-situ' loading created by interference fitting a titanium plug.
Although this strain field is compatible, the method is applicable to a broader class of problems, for example residual strains, as it makes no assumption of compatibility. 
To this end, future work involves the planning of a three-dimensional residual strain experiment.
This could also involve adapting the Gaussian process model to be more suitable for strain fields exhibiting rapid changes or discontinuities.

This is an important step towards the development of strain tomography techniques that can be applied to complex engineered components.
While comparison with neutron diffraction measurements was used for validation, we would argue that these methods are complementary rather than opposing.
These methods are substantially different; neutron diffraction can measure the average strain within a gauge volume at a specific location, whereas tomography methods attempt to reconstruct the entire strain field.
Therefore, we would suggest diffraction measurements are a good choice if a specific region of interest is known, while tomographic methods, such as the one presented in this paper, provide a good alternative if the user wishes to analyse the full field.


Additionally, it was noted that the strain measurements were fewer and of poorer quality than has been previously achieved and that this may have affected the accuracy of the reconstruction. 
Although this issue may be somewhat alleviated by an increase in source power, future work will also investigate full pattern fitting methods \citep{luzin2011residual,sato2013upgrade,sato2017further} which could provide better measurement statistics by analysing multiple Bragg-edges.
Full pattern fitting may also provide a path to extending the method to samples that contain significant texture. 

Finally, future work should also investigate methods for validating the results when other data sets are not available; such as cross validation \citep{devijver1982pattern}.


\section*{Acknowledgements} 
\label{sec:acknowledgements}
This work was supported by the Australian Research Council through a Discovery Project Grant No. DP170102324. Access to the RADEN and KOWARI instruments was made possible through the respective user-access programs of J-PARC and ANSTO (J-PARC Long Term Proposal No. 2017L0101 and ANSTO Program Proposal No. PP6050).
The authors would also like to thank AINSE Limited for providing financial assistance (PGRA) and support to enable work on this project.


\begin{appendix}
\section{Modifications to the Method} 
\label{sec:modifications_to_the_method}
This section provides details on the required modifications to the method presented in \citet{hendriks2019bayesian}. 
This method is modified to adapt the measurement model from one suitable to high-energy X-ray measurements to a model suitable for Bragg-edge neutron transmission measurements, to use individual variances for each each measurement, and to include artificial traction measurements (following \citet{hendriks2018traction}).
This method models the strain field as a Gaussian process;
\begin{equation}
    \bar{\boldsymbol\epsilon}\sim\mathcal{GP}\left(\mathbf{0},\mathbf{K}_{\epsilon\epsilon}(\mathbf{x},\mathbf{x}')\right),
\end{equation}
where $\bar{\boldsymbol\epsilon} = \pmat{\epsilon_{xx} & \epsilon_{yy} & \epsilon_{zz} & \epsilon_{xy} & \epsilon_{xz} & \epsilon_{zz}}^\Transp$ is a vector of the unique components of strain, and $\mathbf{K}_{\epsilon\epsilon}(\mathbf{x},\mathbf{x}')$ is a covariance function designed to ensure all estimated strain fields satisfy equilibrium.

Since both the LRT \eqref{eq:LRT} and the traction measurement model are linear operators, the strain field estimate at location $\mathbf{x}_*$, denoted as $\bar{\boldsymbol\epsilon}_*$, and the measurements are jointly Gaussian;
\begin{equation}\label{eq:joint_distribution}
    \begin{bmatrix}
        \mathbf{y} \\
        \mathbf{T} \\
        \bar{\boldsymbol\epsilon}_*
    \end{bmatrix}
     = \mathcal{N}\left(\begin{bmatrix}
         \mathbf{0} \\ \mathbf{0} \\ \mathbf{0}
     \end{bmatrix},\begin{bmatrix}
         \mathbf{K}_{yy}+\Sigma_n & \mathbf{K}_{Ty}^\Transp & \mathbf{K}_{y\epsilon}^\Transp \\
        \mathbf{K}_{Ty} & \mathbf{K}_{TT}+\sigma_T^2\mathbf{I} & \mathbf{K}_{\epsilon T}^\Transp \\
         \mathbf{K}_{\epsilon I} & \mathbf{K}_{\epsilon T}& \mathbf{K}_{\epsilon\epsilon}
     \end{bmatrix}\right),
\end{equation}
where $\mathbf{y}$ and $\mathbf{T}$ are vectors of all the LRT and traction measurements, respectively.
Here, $\Sigma_n$ is the variance matrix with the variance of each LRT measurement as an entry on the diagonal, $\sigma_t$ is a small variance placed on the traction measurements for numerical stability, and the required modifications are described in the following. 

Within \citep{hendriks2019bayesian} the high-energy X-ray strain measurements were modelled using the line integral
\begin{equation}\label{eq:x_ray_model}
    y = \frac{1}{L}\int_{0}^{L}\hat{\mathbf{\kappa}}^{\Transp} \boldsymbol\epsilon(\hat{\mathbf{n}}s+\mathbf{p})\hat{\boldsymbol{\kappa}}\,\mathrm{d}s + e,
\end{equation}
which differs from the LRT \eqref{eq:LRT} only in that the measurement direction $\hat{\boldsymbol\kappa}$ is a unit vector almost perpendicular to the direction of the ray $\hat{\mathbf{n}}$.
Hence the method is easily adapted by substituting $\hat{\boldsymbol\kappa} = \hat{\mathbf{n}}$, giving the covariance between a strain estimate and the $i^\text{th}$ measurement, $y_i$, as
\begin{equation}
    (\mathbf{K}_{*})_{i} = \frac{1}{L_i}\int\limits_0^{L_i}\mathbf{K}_\epsilon(\mathbf{x}_*,\mathbf{p}_{i}+\hat{\mathbf{n}}_is')\bar{\mathbf{n}}_i^\Transp\,\mathrm{d}s',
\end{equation}
where $\bar{\mathbf{n}} = \pmat{n_x^2 & n_y^2 & n_z^2 & 2n_x n_y & 2n_x n_z & 2n_y n_z}$,.
The covariance between each pair of measurements, $y_i$ and $y_j$ is similarly given by
\begin{equation}
\begin{split}
    (\mathbf{K}_I)_{ij} 
    &= \frac{1}{L_iL_j}\int\limits_0^{L_i}\hspace{-0.8mm}\int\limits_0^{L_j}\bar{\mathbf{n}}_i\mathbf{K}_\epsilon(\mathbf{p}_i+\hat{\mathbf{n}}_is,\mathbf{p}_j+\hat{\mathbf{n}}_js')\bar{\mathbf{n}}_j^\Transp \,\mathrm{d}s'\mathrm{d}s.
\end{split}
\end{equation}

Following the work by \citet{hendriks2018traction}, a traction measurement can be modelled as
\small
\begin{equation}
    T_i = \underbrace{\pmat{n_{\perp 1} & 0 & 0 &  0 & n_{\perp 2} & n_{\perp 3} \\ 0 & n_{\perp 2} & 0 & n_{\perp 1} & 0 & n_{\perp 3}\\
    0 & 0 & n_{\perp 3} & n_{\perp 1} & n_{\perp 2} & 0}\mathbf{C}}_\mathbf{H}
    \bar{\boldsymbol\epsilon}(\mathbf{x}_s),
\end{equation}
\normalsize
where $\mathbf{x}_b$ is a load-free point on the surface with surface normal $\mathbf{n}_\perp = \pmat{n_{\perp 1} & n_{\perp 2} & n_{\perp 3}}$, and $\mathbf{C}$ is the stiffness matrix for isotropic materials from Hooke's law;
\small
\begin{equation}
    \mathbf{C} = \frac{E}{(1+\nu)(1-2\nu)}\pmat{1-\nu & \nu & \nu & 0 & 0 & 0 \\
    \nu & 1-\nu & \nu & 0 & 0 & 0 \\
    \nu & \nu & 1-\nu & 0 & 0 & 0 \\
    0 & 0 & 0 & 1-2\nu & 0 & 0 \\
    0 & 0 & 0 & 0 & 1-2\nu & 0 \\
    0 & 0 & 0 & 0 & 0 & 1-2\nu}.
\end{equation}
\normalsize

Here we will represent the mapping from an estimate of strain to the $k^\text{th}$ traction measurement as $T_k = \mathbf{H}_k\bar{\boldsymbol\epsilon}_*$, which allows the following covariance to be given;
\begin{equation}
\begin{split}
    (\mathbf{K}_{\epsilon T})_k &= \mathbf{K}_{\epsilon\epsilon}\left(\mathbf{x}_*,\mathbf{x}_{s i}\right)\mathbf{H}_k^\Transp, \\
\end{split}
\end{equation}
and the covariance between traction measurement $T_k$ and LRT measurement $y_i$ as
\begin{equation}
\begin{split}
    (\mathbf{K}_{Ty})_{ki} &= \frac{1}{L_i}\int\limits_0^{L_i}\mathbf{H}_k\mathbf{K}_\epsilon(\mathbf{x}_*,\mathbf{p}_{i}+\hat{\mathbf{n}}_is')\bar{\mathbf{n}}_i^\Transp\,\mathrm{d}s'. \\
\end{split}
\end{equation}
Finally, the covariance between a pair of traction measurements $T_k$ and $T_q$ is given by
\begin{equation}
    (\mathbf{K}_{TT})_{qk} = \mathbf{H}_q\mathbf{K}_{\epsilon\epsilon}\left(\mathbf{x}_*,\mathbf{x}_{s i}\right)\mathbf{H}_k^\Transp, \\
\end{equation}

Once the values for the LRT measurements are known, we can give the strain estimates by using conventional Gaussian conditioning. The rest of the implementation details are the same as those in \citet{hendriks2019bayesian}.


\end{appendix}

\bibliography{References}

\end{document}